\documentclass[prl,twocolumn,aps,showpacs,amsmath,amssymb]{revtex4-1}

\usepackage{graphicx}
\usepackage{epstopdf}
\usepackage{dcolumn}
\usepackage{bm}
\usepackage{amsmath}
 
\begin{document}
	
	\title{Competing Orders and Ultrafast Energy Transfer at the Quantum Limit in a Nb$_3$Sn Superconductor Probed by Terahertz Electrodynamics}
	
	\author
	{X. Yang$^{1}$, X. Zhao$^{1}$, C. Vaswani$^{1}$, C. Sundahl$^{2}$, Y. Yao$^{1}$, M.~Mootz$^{3}$, P.~P.~Orth$^{1}$, J. H. Kang$^{2}$,  I.~E.~Perakis$^{3}$, C-Z Wang$^{1}$, K-M Ho$^{1}$, C. B. Eom$^{2}$ and J. Wang$^{1}$}
	
	\affiliation{$^1$Department of Physics and Astronomy and Ames Laboratory-U.S. DOE, Iowa State University, Ames, Iowa 50011, USA. 
		\\$^2$Department of Materials Science and Engineering, University of Wisconsin-Madison, Madison,
		WI 53706, USA.
		\\$^3$Department of Physics, University of Alabama at Birmingham, Birmingham, AL 35294-1170, USA.}
	
	\date{\today}
	
	\begin{abstract}
		We report the low--energy electrodynamics of a moderately clean A15 superconductor (SC) following ultrafast excitation
		to understand and manipulate terahertz (THz) quasi--particle (QP) transport 
		from competing orders.  
		Using 35-fs optical pulses, we observe a non-thermal enhancement in the low frequency conductivity, opposite to that observed for THz pump, 
		which persists up to an additional critical temperature, above the SC one, from an electronic order in the Martensitic normal state. 
		In the SC state, the fluence dependence of pair breaking kinetics
		together with an analytic model provides evidence for a ``one photon--to--one pair'' non-resonant energy transfer during the laser pulse.
		Such initial transfer of photon energy $\hbar\omega_{}$ to QPs at the {\em quantum} limit, set by $2\Delta_{SC}/\hbar\omega$=0.33$\%$, is more than one order of magnitude smaller than in previously studied BCS SCs, which we attribute to strong electron--phonon coupling and possible influence of phonon condensation in A15 SCs. 
	\end{abstract}
	
	\maketitle
	The competition and interference between SC and other co-existing electronic instabilities appear to be universal in quantum materials.
	How to exploit the balance of these orders as a control knob to achieve ultrafast manipulation of materials properties is an outstanding challenge. 
	Answering these questions has been proven important not only in the more sophisticated quantum materials \cite{li2013, Patz2014, Porer} but also in some well-established systems such as A15 superconductors \cite{McMillan,Kataoka, Bhatt,Weber,Sadigh, Freericks}.
	Recently, a strikingly long-lived, gapless quantum phase of quasi-particles with coherent transport is demonstrated by THz quench of a Nb$_3$Sn superconducting gap without heating other degrees of freedom \cite{Xu}. 	
	Nb$_3$Sn, as a paradigmatic A15 compound, exhibits a Martensitic normal state transition above the superconducting one, which has been ascribed to optical phonon condensation (``dimerization'' of Nb atoms) ~\cite{Shirane}, possibly driven by a Van Hove singularity (VHS)-like electronic density-of-states peaked at $\sim$E$_F$ and by strong electron-phonon interaction \cite{Bhatt,McMillan,Weber,CDW,MARKIEWICZ,Sadigh}. 
	The order parameters competing with SC likely exhibit multiple components, both lattice and electronic, e.g., the apparent splitting of the three-fold degenerate $\Gamma_{12}$ band concurrently with an elusive electronic or possibly charge-density-wave(CDW)-like order contribution inferred from tunneling \cite{CDW,CDW2, Freericks} and Raman spectroscopy experiments \cite{Devereaux}.
	As a result, the partial Fermi surface gapping, $\Delta_{W} \gg \Delta_{SC}$ associated with the Martensitic anomaly affects the electronic states near E$_F$ differently from the competing SC order $\Delta_{SC}$  \cite{McMillan}. This opens an opportunity for testing a compelling hypothesis of ultrafast non-thermal manipulation of conductivity in A15 compounds via wavelength-selective pumping.
	
	THz spectroscopy is a powerful tool for {\em quantitative} studies of SC states both in-- and out--of--equilibrium. Arising from energy scales in the vicinity of SC gaps $\Delta_{SC}$ of few meV, terahertz (THz) electrodynamics, characterized by the complex optical conductivity response function $\tilde{\sigma}=\sigma_1(\omega) +i\sigma_2(\omega)$, is a direct measure of both dissipation of QPs and inductivity of SC condensate. 
	Such THz measurements allow access to the key properties of the broken symmetry states, including, e.g., the superfluid/QP density and SC coherence. 
	Prior THz studies of SC samples mostly revealed ``conventional'' features consistent with the deep impurity limit, $\hbar/\tau_{imp} \gg 2\Delta_{SC}$ \cite{Demsar2003PRL,Demsar2003PRL,Beck2011PRL,Shimano2012}. 
	In addition, the spectral-temporal dynamics of the order parameters out-of-equilibrium allow the identification of correlation gaps and co-existing orders.
	The time resolution can follow Cooper pair breaking dynamics
	while the time-resolved THz spectra can track possible spectral weight transfer to the Fermi surface from both the condensate peak $n_s\delta (\omega)$ and possible competing correlation gaps $\Delta_{W} \gg \Delta_{SC}$. 
	These salient features have never been identified in A15 superconductors and the much-needed comparisons are absent in prior THz studies, e.g., in MgB$_2$ and NbN-based superconductors without competing orders \cite{Demsar2003PRL,Beck2011PRL,Shimano2012}. 
		
	In this Letter, we present the fs optically-induced, THz electrodynamics of a moderately clean A15 superconductor Nb$_3$Sn.
	Our results show that the non-equilibrium THz conductivity after fs optical pump excitation ($\sim$1.55eV) gains an additional spectral weight that, strikingly, persists far above the superconducting T$_C$. This is opposite to the case of a THz pump ($\sim$4meV) or increased temperature, which gives rise to a reduced THz conductivity above T$_C$. 	  
	We interpret such pump control of conductivity as non-thermal softening of an electronic correlation gap $\Delta_W$ in the Martensitic phase.
Furthermore, we observe a rapid, SC pair breaking process consistent with strong electron-phonon coupling. This, together with an analytic model, provides evidence for a quantum limit, energy transfer during $\sim$35-fs optical pulses, i.e., one high energy photon breaks only one low energy Cooper pair, with the rest of photon energy exciting phonons rather than creating additional QPs. 
Such ``single quanta", initial transfer of photon energy $\hbar\omega_{}$ to QPs, determined by $2\Delta_{SC}/\hbar\omega$=0.33$\%$, in the very early quantum regime sets an unusual initial condition that determines subsequently {\em ps} pre-bottleneck dynamics and is at least one order of magnitude lower than any previously measured BCS superconductors \cite{Demsar2003PRL, Beck2011PRL}.
	
	A 
	Nb$_3$Sn film 20nm thick was grown by magnetron sputtering on a 1mm Al$_2$O$_3$(R-plane) 
	substrate by co-sputtering of Nb and Sn at high temperatures. 
	%
	The optical pump and THz probe spectroscopy technique is implemented by using three pulses \cite{Luo,Xu}: optical pump E$_{op}$, THz probe E$_{THz}$ by optical rectification, and optical gating pulse at time t$_{gate}$ for electro-optic sampling. 
	The setup was driven by a 1 kHz Ti:Sapphire regenerative amplifier with 35 fs duration at 800 nm center wavelength. 
	
	\begin{figure}[tbp]
		\includegraphics[scale=0.5]{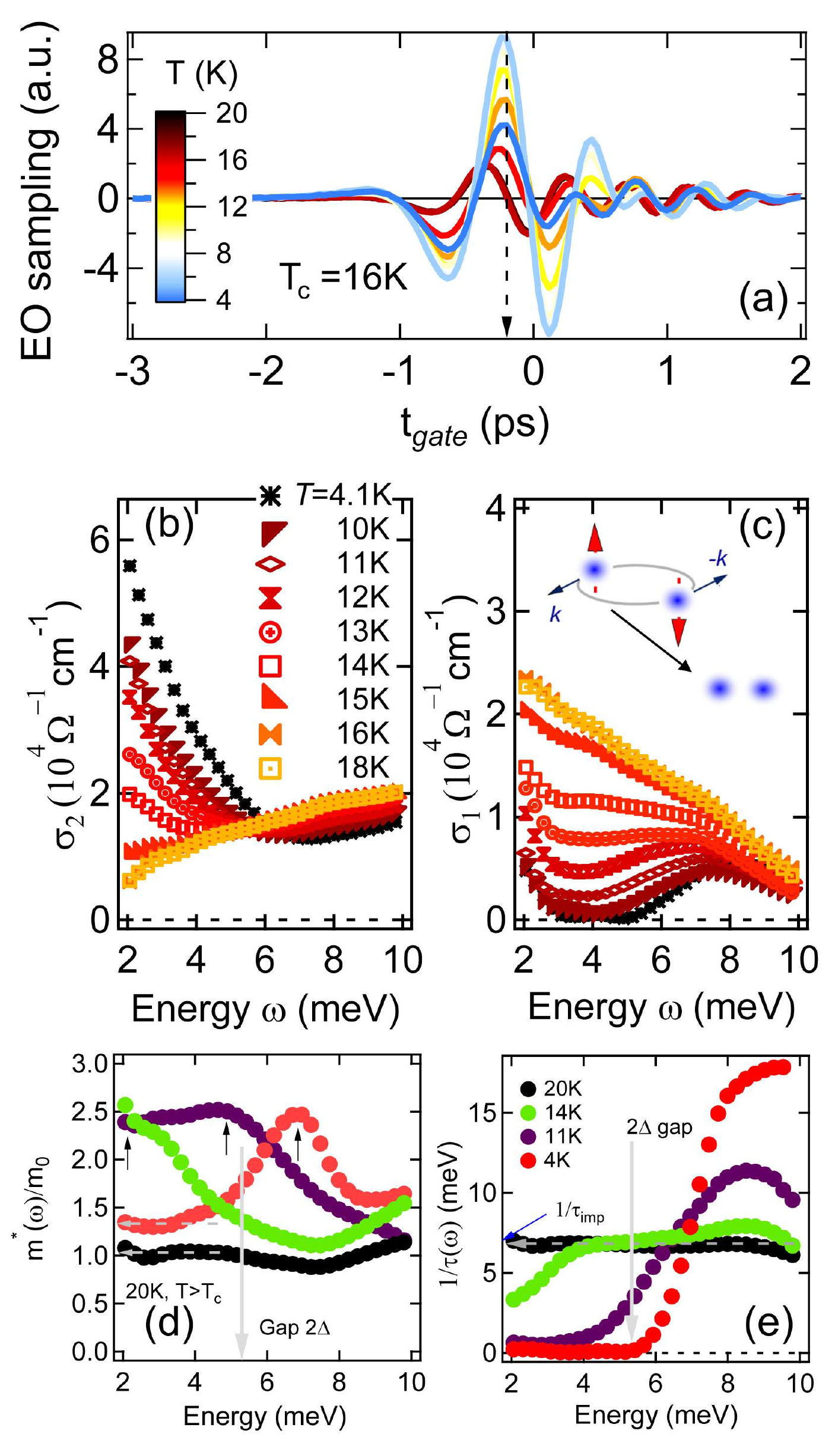}
		\caption{
			(Color online). (a) THz probe transmitted field  E$_{probe}$ as function of gate delay time t$_{gate}$ for the thermal equilibrium state from 4K to 20K. (b), (c) Temperature dependence of imaginary and real parts of the conductivity, $\sigma_1(\omega)$ and $\sigma_2(\omega)$. Inset to (c):  schematic of Cooper pair breaking. 
			(d) Mass renormalization $m^*/m$ and (e) momentum scattering rate $1/\tau$ spectra calculated from $\sigma_2(\omega)$ and $\sigma_1(\omega)$ in (b), (c). Grey solid line denotes $2\Delta_{SC}$ gap at 4.1K. Dashed lines mark the asymptotic $m^*/m$ and $1/\tau$ towards zero frequency.
		}
		\label{Fig1}
	\end{figure}
	
	The equilibrium time-domain THz transmission field and electrodynamics are shown in Figs.~1(a) and 1(b)-1(c), respectively, as a function of temperature. 
	The 4.1\,~K traces exhibit a diverging 1$/\omega$ response in ${\sigma}_2$,  arising from reactive SC condensate, and a dissipationless conductivity, witnessed in $\sigma_1$ below 2$\Delta_{SC}=$5.1\,~meV. A finite $\sigma_1$ peak at the lowest frequencies $<$3\,~meV originates from intraband absorption of the thermally excited Bogoliubons.
	Such conductivity features of SC diminish when approaching T$_c$, as seen in the 10-15\,~K measurements shown in Figs.~2(b)-2(c). As shown in the 16\,~K and 18\,~K traces, the normal state exhibits a Drude response: a gapless $\sigma_1 (\omega)$ and gradually decreasing  $\sigma_2 (\omega)$ at low frequencies. 
	The relatively narrow linewidth of  $\sigma_1 (\omega)$ indicates a much smaller impurity scattering rate, $\hbar/\tau \sim$7 meV, than in previous THz studies \cite{Demsar2003PRL,Beck2011PRL,Shimano2012}.    
	This also leads to more than two orders of magnitude larger $l/\xi$ ratio of mean free path over coherence length in our sample, where $l=v_{F}\tau=$32 nm$\sim$4.5-8$\xi_{exp}$ indicative of a moderately clean SC (supplementary).  
	
	We now extract the optical self energy $\Sigma_{}(\omega, T)$ using an extended Drude model \cite{Carbotte2008}, which provides information complementary to $\widetilde{\sigma}(\omega)$ and more relevant to characterize impurity scattering and correlation. 
	Figs.~1(d) and 1(e) present the complex $\Sigma_{}(\omega, T)$ in terms of the frequency-dependent mass renormalization $m^{*}(\omega)/m_{0}$ and momentum scattering rate $1/\tau_{}(\omega)$, which relate to the real and imaginary parts of the $\Sigma_{}(\omega, T)$, respectively. We emphasize three key observations in the SC state. First, the $1/\tau_{}(\omega)$ spectra in Fig. 1(e) clearly reveal a SC gap opening and suppresses the scattering rate below 2$\Delta_{SC}$ and reduces it to zero at 4.1K. Second, sharp impurity peaks, commonly seen in dirty limit SC samples at 2$\Delta_{SC}$ \cite{Carbotte2008}, are absent in $1/\tau_{}(\omega)$ and replaced by a broad cusp in $m^{*}(\omega)/m_{0}$ above 2$\Delta_{SC}$. Third, $m^{*}(\omega)/m_{0}$ in the SC state as $\omega \rightarrow$0 reflects $n/n_{s}$, i.e., the ratio between the electron density $n$ and the superfluid density $n_s$. Here, $n/n_{s}(4.1K)=m^{*}(\omega=0, 4.1K)/m_{0} \sim$1.34 indicates that $\sim$75$\%$ of the electrons participate in superfluidity, consistent with the superfluid density ratio ($\sim$70$\%$) obtained from the optical sum rule $\int_{0+}^{\infty} (\sigma_1^n(\omega)-\sigma_1^s(\omega))\,d\omega=\frac{\pi}{2}\frac{n_se^2}{m}$. This measured $n_{s}/n$ is $\sim$6 times larger than in superconducting Pb \cite{Carbotte2008}. 
	
	\begin{figure}[tbp]
		\includegraphics[scale=0.35]{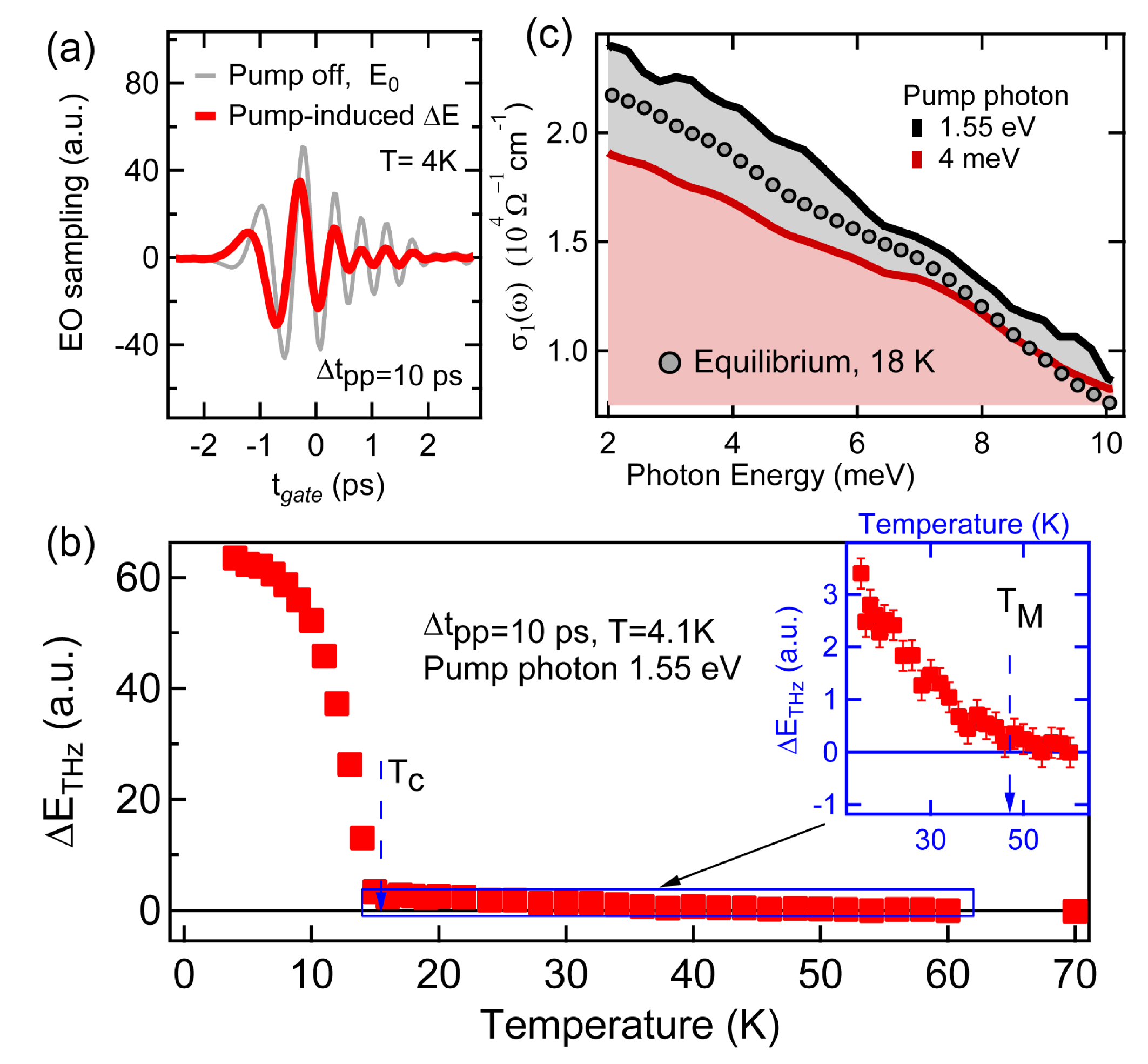}
		\caption{
			(Color online). (a) Transmitted E$_{probe}$ through unpumped Nb$_3$Sn film (gray) and pump induced change $\Delta E_{probe}$ (red). (b) Temperature dependence of peak-to-peak $\Delta E_{probe}$ at 4.02$\mu$J$/cm^2$. $\Delta E_{probe}$ above T$_c$ is magnified in inset to (b) and critical temperature T$_M$ is marked by blue dashed line. (c) $\sigma_1(\omega)$ at 10 ps after 1.55eV (black), 4meV (red) photo-excitation compared to equilibrium (gray circle) at 18K.}
		\label{Fig2}
	\end{figure}
	
	Fig.2 demonstrates the ultrafast study of the Martensitic phase in the normal state of Nb$_3$Sn. The measured THz fields are shown in the time domain in Fig. 2(a), including the pump induced change, $\Delta E_\text{}(t)$, at a fixed pump-probe delay $\Delta \tau_{pp}$=10 ps (red) and the transmitted field through the unexcited sample, $E_\text{0}(t)$ (gray). 
	In Fig.~2(b) we record the photoinduced THz field peak-to-peak amplitude as a function of temperature from 4K to 70K. Two transitions are visible. As expected for a SC, the photoinduced signal drops significantly at T$_C \sim$16K. Unlike in a conventional SC, however, the signal persists into the normal state and completely vanishes only at a much higher temperature T$_M \sim$47\,~K (inset). 
	Intriguingly, the latter transition in the THz field amplitude 
	coincides with the Martensitic anomaly associated with the structural-electronic instabilities \cite{Bhatt,McMillan,Weber,CDW,MARKIEWICZ,Sadigh,Shirane}.  
	Fig. 2(b) reveals such additional order parameter with a critical temperature T$_M$.
	
	Importantly, the competing T$_M$ order above SC allows ultrafast {\em non-thermal} control of the low-energy THz conductivity response by tuning the pump between the optical and THz frequency range above or below the correlation gap $\Delta_{W}$. We start with the normal state at 18K slightly above T$_C$. The non-equilibrium $\sigma_1(\omega)$ data is shown in Fig. 2(c) for 1.55eV (optical, black) and 4meV (THz, red) pump photon energy. After 1.55eV pump excitation (black), the low frequency conductivity $\sigma_1(\omega)$ gains an additional spectral weight over its equilibrium (no pump) values (gray circles), which is responsible for the non-vanishing, pump-induced signals $\Delta E_{THz}$ in the normal state below T$_M$ shown in Fig. 2(b).  This pump-induced enhancement is consistent with non-thermal softening of the correlation gap that develops at the T$_M$ transition from $\Gamma_{12}$ phonon condensation (dimerization) and/or electronic VHSs, by optical excitation with $\hbar\omega_{op}\gg\Delta_{W}$.  Such softening gives rise to spectral weight transfer to the Fermi surface from high energies above $\Delta_{W}$. Note that laser-induced heating will decrease the THz spectral weight after the pump in Nb$_3$Sn due to very subtle partial gapping of the Fermi surface, opposite to typical CDW materials (supplementary). 
	Intriguingly, by changing the pump photon energy to 4 meV, i.e., $\hbar\omega_{THz}\ll\Delta_{W}$, we observe pump-reduced instead of pump-enhanced conductivity (red).  
	This opposite behavior 
	indicates that photoexcitation at sufficiently low frequencies fails to strongly quench the $\Delta_{W}$ gap and, instead, depletes the Fermi sea portion with a non-thermal, threshold behavior (supplementary).  
In all cases, the observed pump-wavelength-dependent beavior represents a direct evidence for ultrafast non-thermal control of the THz conductivity by tuning pump photoexcitation above (correlation gap melting) and below (Fermi sea partial depletion) $\Delta_{W}$. 
	Note that the unusual pump wavelength dependence is absent in previously studied density wave materials \cite{Kim2015,Patz2014,Porer} and BCS SCs without co-existing orders \cite{Demsar2003PRL,Beck2011PRL, Shimano2012}. The distinct non-thermal photoexcitation QP control of conductivity provides evidence for an additional electronic instability associated with the Martensitic anomaly beyond the conventional structural one.  

	\begin{figure}[tbp]
		\includegraphics[scale=0.45]{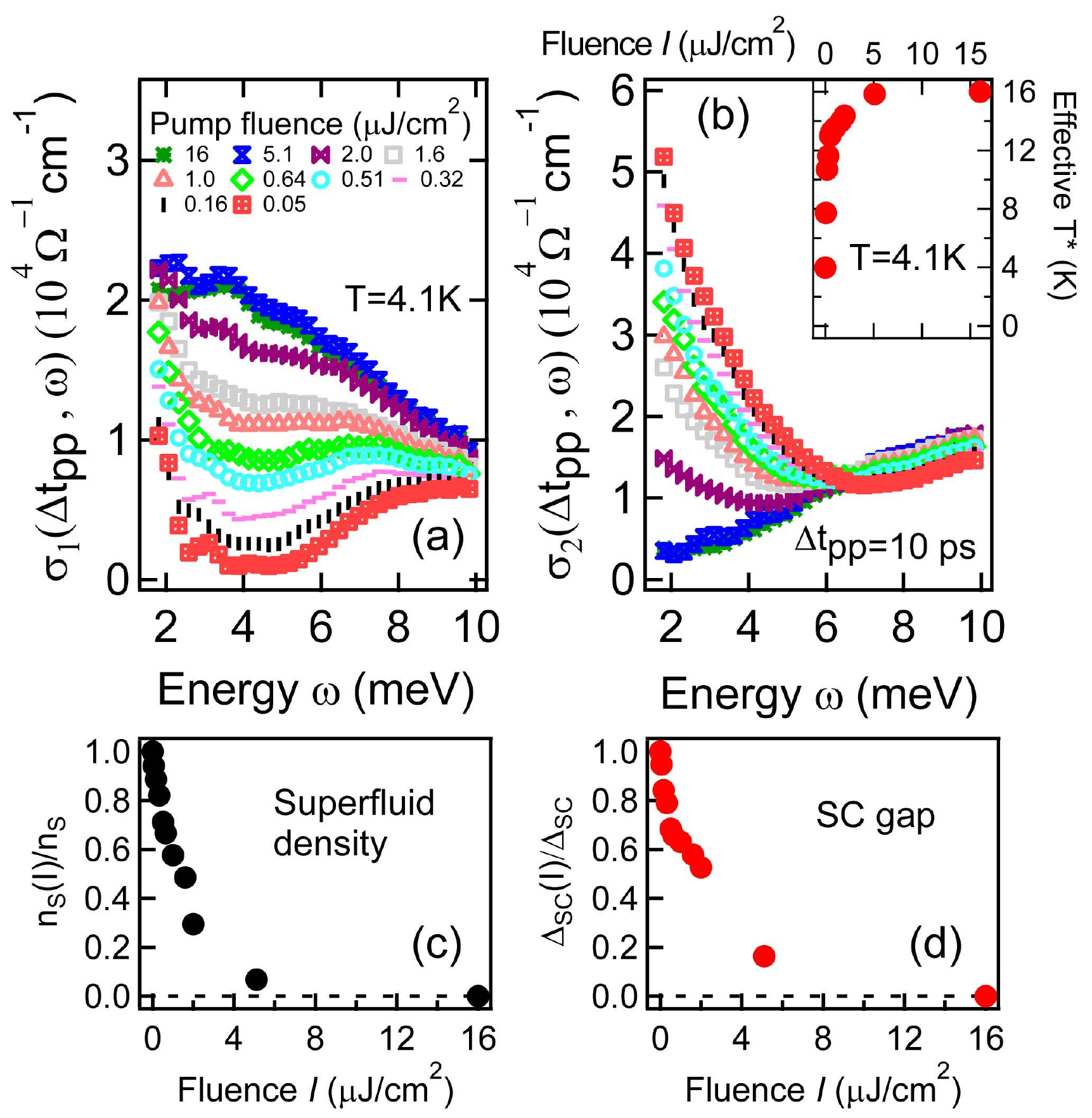}
		\caption{
			(color online). (a), (b) Non-equilibrium  $\sigma_1(\omega)$ and $\sigma_2(\omega)$ at pump-probe delay t$_{pp}$=10ps for  optical-pump fluences 0.05-16$\mu J/cm^2$. Inset to (b) shows effective temperature T$^*$ at various fluences. (c), (d) Fluence dependence of superfluid density n$_s$ and SC gap $\Delta_{SC}$ extracted from (a)-(b). 
		}
		\label{Fig3}
	\end{figure}
	
	Next we turn our attention to the non-equilibrium Cooper pair breaking (CPB) responses in the superconducting state after fs optical excitation. 
	For $\hbar\omega_{op}>> 2\Delta_{SC}$, CPB processes can be driven by multiple interactions between the condensate and photoexcited QPs or with high frequency phonons (HFPs). 
	Previous works have shown that the majority of the absorbed photon energy subsequently transfers to the phonon reservoir during the fs excitation and then continues to break Cooper pairs \cite{Demsar2003PRL, Beck2011PRL}. 
	Figs. 3(a) and 3(b) plot the non-equilibrium THz conductivity $\sigma_{1}(\omega)$ and $\sigma_{2}(\omega)$ of Nb$_3$Sn, for various fluences of 1.55eV pump photoexcitation at $T$=4.1 K. We observe very similar 
	spectral shapes to those seen at various temperatures in equilibrium (Fig. 1). 
	Photo-induced QPs gradually close the SC gap $2\Delta_{SC}$. 
	At the same time, the low frequency $1/\omega$ divergence in $\sigma_{2}$ diminishes with increasing pump fluence I$_{pump}$. SC features disappear simultaneously above 4$\mu$J$/cm^2$.  The thermalized gap $\Delta_{SC}(I_{pump})$ and the superfluid density n$_s$, readily obtained from transient THz spectra shown in Figs. 3(c) and 3(d), quickly diminish as I$_{pump}$ approaches the same fluence. 
	An elevated elecron/lattice transient temperature T$^*$ established after the pump can be extracted by fitting the conductivity data. As shown in the inset of Fig. 3(b), T$^*\rightarrow$T$_c$ is clearly visible at the quenching fluence $\sim$4$\mu$J$/cm^2$. Therefore, conductivity at $\Delta t_{pp}$=10 ps for T$<$T$_C$ is consistent with previous T$^*$ model of non-equilibrium superconductivity \cite{owen,Carbotte2008,Kabanov2005PRL}. 
	\begin{figure}[tbp]
		\includegraphics[scale=0.45]{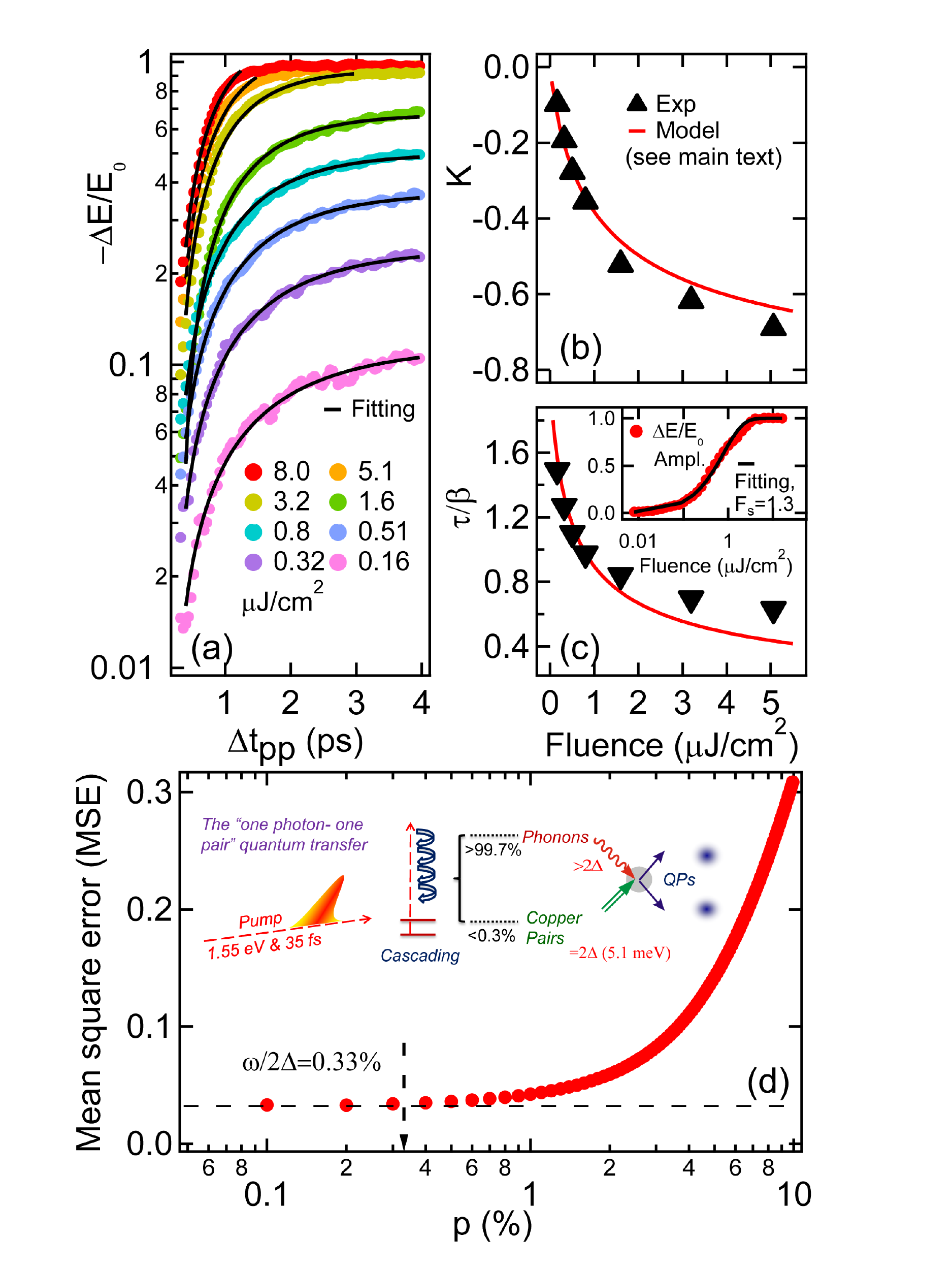}
		\caption{(color online). (a) Pump-probe dynamics measured in experiment (dots) and fitted by RT model (black line). (b), (c) Fluence dependence of RT model  parameters K and $\tau/\beta$ (black triangle) and fitting curve (red line). (d) Fitting MSE vs. QP energy absorption percentage p. Inset to (c) shows fluence dependence of pump induced $\Delta E$ fitted by a saturation curve $(1-exp(-I/F_s))$. Inset to (d) shows the schematics of microscopic CPB by 1.55eV photon.}
		\label{Fig4}
	\end{figure}
	
	We track the early time, pair breaking dynamics prior to the establishment of a T$^*$ quasi-equilibrium temperature regime due to scattering of condensate with HFPs and hot QPs. 
	The experimentally--observed ultrafast THz signals for various pump fluences, which reflect the photoexcited QP density $n$(t), are presented in Fig. 4(a). They show faster CPB with increasing pump fluence during  the first 4 ps. In order to reveal the microscopic energy transfer among various reservoirs, we model the CPB kinetics based on the widely used, Rothwarf-Taylor (RT) model \cite{Rothwarf1967PRL} that is extensively discussed in \cite{Kabanov2005PRL}. This model is expected to be valid at later ps timescales following the "initial condition" created during the very early 10's fs regime of ultrafast dynamics dominated by electron-phonon coherences and by quasi-particle and phonon nonthermal populations created during the laser pulse. The dynamics after this initial regime is characterized by QP and HFP densities, $n$(t) and $N$(t), described by two coupled differential equations \cite{Rothwarf1967PRL, Kabanov2005PRL, Demsar2003PRL, Beck2011PRL}. 
	The rise of $n$(t) in time originates from the pre-bottleneck, CPB process preceding the QP relaxation, which can be described by the analytical expression \cite{Kabanov2005PRL}:
	\begin{eqnarray}  \label{equ5}
	\begin{aligned}
	n(t)=\dfrac{\beta}{R} [-\dfrac{1}{4}-\dfrac{1}{2\tau}+\dfrac{1}{\tau}\dfrac{1}{1-Kexp(-t\beta/\tau)}],
	\end{aligned}
	\end{eqnarray}
	where K and $\tau$ are dimensionless parameters determined by the initial conditions: $K=((\tau/2)/(4Rn_0/\beta+1)-1)/((\tau/2)/(4Rn_0/\beta+1)+1))$and $\dfrac{1}{\tau}=\sqrt{1/4+2R/\beta(n_0+2N_0)}$.  
	Here, $\beta$ is the CPB probability by absorption of HFP and $R$ is the bare QP bi-molecular decay rate. n$_0$ and N$_0$ are the initial QP and HFP densities immediately after fs photoexcitation.
	$\beta$ and $R$ are fluence independent parameters under weak excitation limit, when $n_0$ is much smaller than the material--dependent value $\beta/R$. 
	Fig. 4(a) presents the best fits, which show a very good agreement with the data. The fitting parameters $\tau/\beta$ and K are extracted as the function of fluence and plotted in Figs. 4(b) and (c) respectively. 
	
	Further quantitative information can be obtained by fitting $\tau/\beta$ and K versus the absorbed energy density $\Omega$.  Here $\Omega$ at $I_{q}$=4.02$\mu$J$/cm^2$ is equal to the SC condensate energy U=4757mJ/mol \cite{Kim2015}. 
	Given the lack of direct information about the very early 10's fs quantum regime during the laser pulse, we treat the initial condition phenomenologically, which can be extracted from the measured {\em ps} dynamics and a rigorous error analysis.
	Denoting the portion of the absorbed energy that initially goes into QP excitation as {\em p}, we have n$_0$=p$\Omega/\Delta$ and N$_0$=(1-p)$\Omega/2\Delta$ created by fs photoexcitation.
	The best fit to the extracted $\tau/\beta$ and K data, obtained by minimizing the mean-square error (MSE) of the parameter set $\left \{ p, R, \beta  \right \}$, is achieved for p=0.2$\pm$0.1$\%$, which gives the values of $\beta^{-1}$ = 1.0 $\pm$ 0.1 ps and R = 105.5 $\pm$ 10 ps$^{-1}$ unit cell$^{-1}$.  
	We further plot the MSE of the above fitting as function of {\em p} in Fig. 4d by only fitting $\left \{R, \beta  \right \}$ for each fixed p (supplementary). 
	Intriguingly, a strong deviation from the minimum error starts at a very small p$\sim$0.33$\%$ that coincides with $2\Delta_{SC}/\hbar\omega$, as marked (dashed arrow) in Fig. 4(d). 
	The merely 0.33$\%$ portion of absorbed photon energy $\hbar\omega_{}$ that excites QPs indicates a quantum limit of the energy transfer process during photoexcitation of the A15 system. This unusually small parameter value obtained from the standard fitting implies that one high energy photon, $\hbar\omega$=1.55 eV, breaks only one pair, $2\Delta_{SC}=$5.1 meV, with simultaneous excitation of phonon populations during the pulse and initial QP cascading (inset, Fig. 4(d)).   
	The good agreement between data and simulations, seen in Fig. 4, validate this claim, which implies that strong electron-phonon couplings dominate in A15 SCs. Here phonon condensation from Martensitic order already is a part of the SC electronic order which can potentially be much more efficiently excited during the pulse that differentiates the A15 from other SCs. 
	The initial photon energy transfer to QPs in Nb$_3$Sn is at least one order of magnitude smaller than in other SCs: 0.33$\%$ here vs. MgB$_2$ (p$\sim$6$\%$) \cite{Demsar2003PRL} and NbN (p$\sim$9$\%$) \cite{Beck2011PRL}. 
	
The applicability of the analytic RT model is
well justified by the following condition satisfied by our data:n$_0={p\Omega}/{\Delta}\simeq 1\times10^{-4}\ll{\beta}/{R}=1\times10^{-2}$, where $\beta/R=\dfrac{N(0)^2\pi\omega_D^3}{18v\Delta _{SC}}$\cite{Kabanov2005PRL}. 
Here $\omega_D$ is the phonon cutoff frequency, N(0) is the electronic density of states per unit cell at the Fermi level and $v$ is the number of atoms per unit cell. We used  the values $v$=8, $\Delta_{SC}$=2.55 meV, N(0)=11.4 states(spin cell eV)$^{-1}$ and $\omega_D$=6.9 THz. 
In addition, the value of electron-phonon coupling constant $\lambda$ can be determined from the relation $R=\dfrac{8\pi v \lambda\Delta_{SC}^2}{\hbar N(0)\omega_D^2}.$  We obtain $\lambda\approx$ 2.0, which agrees very well with previous estimates of $\lambda \approx$1.8$\pm$0.15 \cite{Wolf1980PRB} and is 2 times larger than in the previously studied NbN. In addition, a much higher phonon-pair scattering probability $\beta \sim$ 1ps$^{-1}$ is seen in Nb$_3$Sn as compared to MgB$_2$ ($\beta$=1/15 ps$^{-1}$) \cite{Demsar2003PRL} and NbN ($\beta$=1/6 ps$^{-1}$) \cite{Beck2011PRL}. These are consistent with our observation of much faster CPB dynamics in Fig. 4(a), and a minimal initial energy transfer to QPs (p) in Fig. 4(d). 
The enhanced $\beta$ is due in part to a much larger density of states $N(0)/v$ resulting from the bandstructure in Nb$_3$Sn (1.425) as compared to NbN (0.44) and MgB$_2$ (0.23) (supplementary). 

In summary, the comprehensive study of THz electrodynamics of Nb$_3$Sn identifies a competing electronic order in the normal state below the Martensitic anomaly. We demonstrate ultrafast non-thermal control of the THz conductivity in the Martensitic normal state by tuning the pump photon energy. 
In the SC state, we reveal a ``one photon--one pair" quantum energy transfer during initial 10's of fs timescales, the only known example, which we attribute in part to 
strong electron--phonon coupling and optical phonon condensation in the ground state. The distinct ultrafast THz electrodynamics of the model A15 compound with electron-phonon complex order and its differences from BCS/density wave systems offer perspectives to manipulate competing orders in other materials \cite{new} and probe the very early, {\em fs} pair breaking dynamics in the quantum nonthermal regime.  

Work at Iowa State University was supported by the Army Research office under award W911NF-15-1-0135 (THz study). Work at Wisconsin (sample growth and basic characterizations) was supported by the DOE Office of Basic Energy Sciences under award number DE-FG02-06ER46327. Work at the University of Alabama, Birmingham was supported by start—up funds. 

\end{document}